%% file: main.tex
\documentclass{article}
\usepackage{spconf,amsmath,graphicx,hyperref,kotex,xcolor}
\usepackage{url}
\usepackage{amssymb}
\usepackage{booktabs}
\usepackage{multirow}
\usepackage{arydshln}
\usepackage{tabularx}
\usepackage{array}
\usepackage[table]{xcolor}

\title{Speaking Longer Than Learned: Training-Free Long-Form Generation for Full-Stream Text-to-Speech}

\name{Jeseung Lim$^{1}$, Jiseong Woo$^{2}$, Heeseung Kim$^{2,\dagger}$\thanks{$\dagger$ Corresponding Author.}}
\address{
$^{1}$Department of Physics, University of Seoul, Seoul, Republic of Korea\\
$^{2}$Department of AI, University of Seoul, Seoul, Republic of Korea
}

\begin{document}
\maketitle
\begin{abstract}
Recent full-stream text-to-speech (TTS) models enable low-latency personalized synthesis as input text arrives.
However, sustaining generation beyond the lengths seen during training without additional training remains largely unexplored.
We systematically analyze VoXtream2, which extends generation by periodically discarding past context while retaining the acoustic prompt.
We find that this strategy does not prevent degradation: pronunciation accuracy deteriorates near the end of the training range due to the range of positions in which the model operates rather than the loss of past context, while speaker similarity gradually declines throughout generation.
Based on these findings, we propose two training-free modifications that reduce the attention window and strengthen attention to the acoustic prompt.
Experiments show that these modifications preserve pronunciation accuracy and speaker similarity throughout long-form generation while retaining the naturalness of the original model.
Ablations further show complementary effects of the two modifications.\footnote{Project page: \url{https://voxtend.github.io}}
\end{abstract}
\begin{keywords}
streaming text-to-speech (TTS), long-form TTS, training-free length extrapolation, attention bias
\end{keywords}

\input{sections/1_introduction}
\input{sections/2_method}
\input{sections/3_experiments}
\input{sections/4_conclusion}

\section{Compliance with Ethical Standards}
The mean opinion score (MOS) evaluation was conducted via an online survey platform with participant consent, and no sensitive personal information was collected.

\section{Acknowledgments}
This work was supported by the Korea Government (MSIT) through the National Research Foundation of Korea (NRF) (No. RS-2022-NR068754, NO. RS-2026-25601142) and by the supercomputing resources provided by the Urban Big Data and AI Institute of the University of Seoul (\url{http://ubai.uos.ac.kr}).

We used AI (LLM) to assist with language editing and with writing code for the demo page.

\bibliographystyle{IEEEbib}
\bibliography{main}

\end{document}

%% file: sections/1_introduction.tex
\section{Introduction}

Recent text-to-speech (TTS) systems support zero-shot voice cloning and expressive speech generation~\cite{du2024cosyvoice2scalablestreaming}, while growing demand for spoken conversational systems has increased interest in streaming TTS.
Recent approaches enable synthesis before the complete input text is available through interleaved speech-text modeling~\cite{bai2025speakstreamstreamingtexttospeechinterleaved}, direct autoregressive generation from incoming text~\cite{torgashov2026voxtream2fullstreamttsdynamic}, or streaming acoustic decoding~\cite{du2024cosyvoice2scalablestreaming}.
These advances have made low-latency TTS increasingly practical for real-time conversational systems.

Beyond short utterances, many applications require long-form synthesis over multiple sentences or paragraphs~\cite{pmlr-v267-park25k}.
Concatenating independently synthesized segments can introduce speaker inconsistencies and boundary artifacts~\cite{ghosh2026magpiettslfinferencetimelongformspeech}, while long-form paired speech data remains scarce.
Consequently, synthesis beyond the sequence lengths observed during training becomes a length extrapolation problem.
Similar challenges have been studied in large language models (LLMs) through additional long-context training or training-free context management~\cite{xiao2024efficient}.
StreamingLLM~\cite{xiao2024efficient} retains a few initial attention sink tokens together with a bounded window of recent tokens to enable inference beyond the training context.

Extending generation beyond the sequence lengths seen during training has also been explored in long-form TTS.
Approaches based on additional training improve length extrapolation through positional modeling or alignment,
while others explicitly train on long-duration speech for long-form generation~\cite{pmlr-v267-park25k}.
Training-free approaches instead avoid additional long-form training; MagpieTTS-LF~\cite{ghosh2026magpiettslfinferencetimelongformspeech} extends TTS models through stateful chunk generation and attention priors.
More recently, long-form synthesis with streaming text input has also been explored, although existing approaches require additional training to operate with a bounded streaming context~\cite{liu2026prosodicboundaryawarestreaminggeneration}.
To the best of our knowledge, training-free length extrapolation for a pretrained full-stream TTS model under continuous long-form generation has not been explored.

In this work, we tackle training-free length extrapolation in continuous long-form generation using VoXtream2~\cite{torgashov2026voxtream2fullstreamttsdynamic}, whose official implementation extends generation beyond the training length by periodically resetting the key-value cache while retaining the acoustic prompt and recent frames, following the attention sink concept~\cite{xiao2024efficient}.
However, this mechanism alone cannot sustain long-form generation: pronunciation accuracy degrades near the end of the training range, while speaker similarity declines gradually throughout the utterance.
We find that the former stems from the positions at which the model operates rather than from the cache reset itself, while the latter persists even when these positions are avoided.
Based on these findings, we shrink the attention window to keep the model within positions encountered during training and bias attention toward the acoustic prompt, the only intact explicit speaker reference in the context.

We first show that the loss of speaker similarity is not unique to VoXtream2, but also occurs in open-source TTS models trained on long-form speech.
Applied to VoXtream2, our methods preserve both pronunciation accuracy and speaker similarity throughout long-form generation with streaming text input.
Human evaluation confirms comparable naturalness, while ablations on attention window size and bias design show complementary effects on long-form stability.

%% file: sections/2_method.tex
\section{Method}

\subsection{VoXtream2}

VoXtream2~\cite{torgashov2026voxtream2fullstreamttsdynamic} is a full-stream TTS model conditioned on reference speech, which is encoded into Mimi codec tokens~\cite{defossez2024moshispeechtextfoundationmodel} and serves as an acoustic prompt.
It consists of a Phoneme Transformer (PT), a Temporal Transformer (TT), and a Depth Transformer (DT).
PT incrementally encodes phonemes, TT autoregressively predicts semantic and duration tokens from these embeddings and previous audio tokens, and DT generates the remaining codebooks.
Each duration token combines a shift state $s\in\{0,1,2\}$, indicating the base phoneme advancement, with a binary phoneme-count state $u\in\{0,1\}$, where $u=0$ and $u=1$ correspond to assigning one and two phonemes to the current frame.
The effective phoneme advancement per frame is $s+u\in\{0,1,2,3\}$.
VoXtream2 is trained on random $50$s crops ($625$ frames at $12.5$Hz).

During long-form generation, the key-value (KV) cache keeps growing and must be bounded.
Streaming\allowbreak LLM~\cite{xiao2024efficient} showed that a bounded cache can remain stable by retaining the first few tokens, known as attention sinks, together with the most recent ones.
The official VoXtream2 implementation follows a similar idea in its \texttt{SinkAttention} module.
TT indexes cached frames by local position and holds at most $W$ frames, where \texttt{audio\_window\_size} sets $W=625$ by default.
Once the local position reaches $W$, a cache reset retains the $P$ acoustic prompt frames and the $\tau=\lfloor W/2 \rfloor$ recent generated frames while discarding the rest.
The frames are re-indexed from $0$ and their KV states recomputed, so generation resumes at local position $P+\tau$.
The local position then climbs back to $W$, triggering the next reset after $W-P-\tau$ frames and cycling between $P+\tau$ and $W$.

\subsection{Analysis of Long-Form Generation Degradation}
\label{subsec:longform}

To examine long-form zero-shot TTS beyond the training length, we use LibriSpeech-Long~\cite{pmlr-v267-park25k}, whose recordings last up to four minutes.
We use the $281$ recordings of its \texttt{dev-clean} split, totaling $15.89$ hours.
For each recording, the opening speech segment ending at a sentence boundary serves as the acoustic prompt, with an average duration of $7.88$s, and the remaining transcript serves as the target text.

\begin{figure}[t]
  \centering
  \includegraphics[width=0.78\columnwidth]{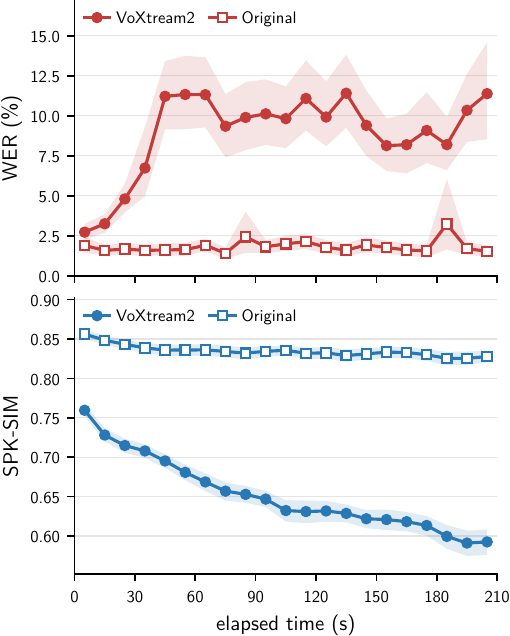}
    \caption{Segment-wise WER and SPK-SIM of VoXtream2, together with the corresponding ground truth speech.}  
    \label{fig:baseline}
  \vskip -14pt
\end{figure}

We measure pronunciation accuracy using word error rate (WER) over $10$s windows, pooling reference words and errors across utterances within each window.
Generated speech is transcribed with \textit{whisper-large-v3}~\cite{pmlr-v202-radford23a}, and its words are time-stamped with \textit{Qwen3-ForcedAligner-0.6B}~\cite{shi2026qwen3asrtechnicalreport}.
After edit-distance alignment to the target text, each error is assigned to the window containing its word, with deletions placed between neighboring words.
SPK-SIM is the cosine similarity between speaker embeddings of generated speech and the acoustic prompt, extracted with ECAPA-TDNN~\cite{desplanques20_interspeech} over the same windows.
As shown in Figure~\ref{fig:baseline}, the cache reset does not prevent two distinct degradation patterns.
WER starts at $2.74\%$ in the first $10$s window and rises more sharply near the training length, reaching $11.23\%$ at $40$--$50$s and remaining high, while SPK-SIM decreases steadily from $0.760$ to $0.695$ over the same interval and continues to decline.

\subsubsection{Position, Not Cache Reset}
\label{sec:position}

Training crops are $625$ frames long, so TT never undergoes a cache reset during training and observes local positions above roughly $600$ only near the end of a crop, when little of the transcript remains.
With the default $W=625$, however, the local position repeatedly enters this region immediately before each cache reset during long-form generation.
The sharp WER increase in Figure~\ref{fig:baseline} may therefore arise either from position-dependent behavior near the end of the training range or from the cache reset itself.
To distinguish these effects, we vary $W$ at inference, moving the reset position while leaving the end of the training range at position $625$ unchanged.

\begin{figure}[t]
  \centering
  \includegraphics[width=0.78\columnwidth]{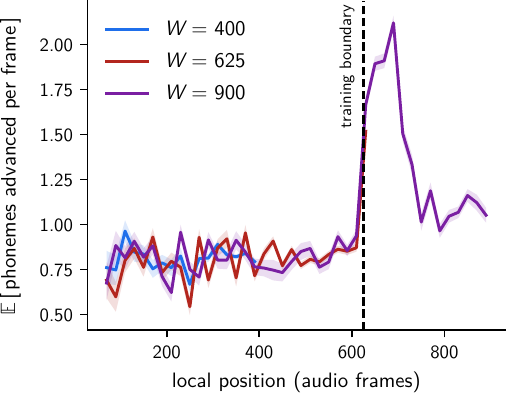}
\caption{Expected number of phonemes advanced per frame versus position for $W=400$, $625$, and $900$. Shaded bands show standard errors over $8$ randomly selected utterances.}
  \label{fig:shift}
  \vskip -14pt
\end{figure}

Figure~\ref{fig:shift} shows that the phoneme advancement rate remains near $0.8$ below position $600$ but rises sharply toward position $625$, reaching $1.9\times$ and $2.6\times$ this rate for $W=625$ and $900$, respectively.
For $W=900$, the increase still occurs near position $625$, although the reset is delayed by $275$ frames, showing that degradation follows local position rather than cache reset.
With $W=900$, generated audio is only $70\%$ as long as ground truth, while deletion errors increase from $0.7\%$ at $W=625$ to $16.6\%$, indicating that the model advances too quickly and skips target words.
WER rises from $7.3\%$ to $49.8\%$ across position $625$, but changes only from $47.8\%$ to $44.2\%$ around the later cache reset.
This motivates our first intervention: restricting $W$ avoids problematic end-of-training positions while retaining recent context.

\input{tables/Table1}

\subsubsection{Attending to the Prompt}
\label{sec:speaker}

Unlike the sharp increase in pronunciation errors above, SPK-SIM declines steadily throughout generation (Figure~\ref{fig:baseline}).
Reducing $W$ does not remove this trend but slows it, from $-0.082$ per $100$s with $W=625$ to $-0.054$ with $W=400$.
Nor does the decline track reset frequency: with $W=400$, resets are more frequent than with $W=625$, yet SPK-SIM degrades more slowly.
We therefore examine whether the acoustic prompt remains important throughout generation.

\begin{figure}[t]
  \centering
  \includegraphics[width=0.78\columnwidth]{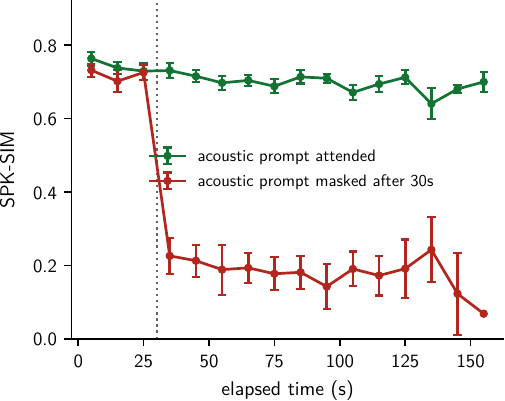}
\caption{SPK-SIM over time with $W=400$ when the acoustic prompt is attended and when it is masked after $30$s. Error bars show standard errors over $8$ randomly selected utterances.}
  \label{fig:speaker}
  \vskip -14pt
\end{figure}

To test this hypothesis, after 30s, we prevent the model from attending to the acoustic prompt keys while retaining the generated-history cache.
With $W=400$, the cache contains at least the most recent $\tau=\lfloor W/2 \rfloor=200$ generated frames, corresponding to $16$s of speech.
Despite this context, SPK-SIM drops from $0.73$ to $0.23$ in the next $10$s window and remains below $0.25$ thereafter (Figure~\ref{fig:speaker}).
This indicates that the acoustic prompt remains essential throughout generation, but its $P$ frames remain fixed while generated frames continue to accumulate in the cache.

We therefore strengthen attention to the acoustic prompt by adding a constant bias $b$ to the pre-softmax attention logits of its keys~\cite{guardieiro2026instructionfollowingprincipledboosting}.
For an attention logit $L_{ij}$ from query $i$ to key $j$, we use $L'_{ij}=L_{ij}+b$ for prompt keys and leave it unchanged otherwise.
The bias is applied to the acoustic prompt keys across all attention heads and layers of TT, requiring neither retraining nor architectural changes.
We also compare with SpotLight~\cite{venkateswaran-contractor-2026-spotlight}, which adaptively adds $\log(\psi_{\mathrm{target}}/\psi_{\mathrm{current}})$ to prompt-key logits when the prompt attention mass $\psi_{\mathrm{current}}$ falls below the target $\psi_{\mathrm{target}}$.
Section~\ref{subsubsec:ablation} compares constant and adaptive prompt biases.

%% file: tables/Table1.tex
\begin{table*}[t]
\centering
\caption{
Overall and segment-wise performance on long-form speech synthesis.
Lower WER (\%) and higher SPK-SIM and MOS indicate better performance.
MOS is reported as mean $\pm$ 95\% confidence interval on a 5-point scale.
}
\label{tab:main_results}

\vspace{1mm}

\small
\setlength{\tabcolsep}{0.8pt}
\renewcommand{\arraystretch}{1.00}

\begin{tabular*}{\textwidth}{
@{}
>{\raggedright\arraybackslash}p{4.0cm}
@{\hspace{4pt}}
l
@{\extracolsep{\fill}}
cccccccc
@{\hspace{5pt}}
c
@{}
}
\toprule

\multirow{3}{*}{\textbf{Model}} &
\multirow{3}{*}{\textbf{Metric}} &
\multicolumn{8}{c}{\textbf{Objective Metrics}} &
\multicolumn{1}{c}{\textbf{Subjective}} \\

\cmidrule(lr){3-10}
\cmidrule(l){11-11}

&
&
\multirow{2}{*}{\textbf{Overall}} &
\multicolumn{7}{c}{\textbf{Segment (s)}} &
\multirow{2}{*}{\textbf{MOS} $\uparrow$} \\

\cmidrule(lr){4-10}

&
&
&
\textbf{0--30} &
\textbf{30--60} &
\textbf{60--90} &
\textbf{90--120} &
\textbf{120--150} &
\textbf{150--180} &
\textbf{180--210} &
\\

\midrule

\multirow{2}{*}{Ground Truth (GT)}
& WER $\downarrow$
& 1.86
& 1.73 & 1.62 & 1.93 & 1.99 & 1.78 & 1.66 & 2.18
& \multirow{2}{*}{$4.39 \pm 0.05$} \\

& SPK-SIM $\uparrow$
& 0.832
& 0.849 & 0.837 & 0.834 & 0.834 & 0.831 & 0.832 & 0.826
& \\

\noalign{\vskip 0.6pt}
\cdashline{1-11}[0.3pt/1.5pt]
\noalign{\vskip 0.6pt}

\multirow{2}{*}{GT + Mimi Codec~\cite{defossez2024moshispeechtextfoundationmodel}}
& WER $\downarrow$
& 1.95
& 1.80 & 1.60 & 1.73 & 2.03 & 2.49 & 1.75 & 1.89
& \multirow{2}{*}{$4.10 \pm 0.07$} \\

& SPK-SIM $\uparrow$
& 0.754
& 0.771 & 0.760 & 0.756 & 0.757 & 0.753 & 0.754 & 0.748
& \\

\specialrule{0.8pt}{1pt}{1pt}

\multirow{2}{*}{Qwen3-TTS-1.7B-Base~\cite{hu2026qwen3ttstechnicalreport}}
& WER $\downarrow$
& 1.59
& 1.31 & 1.38 & 1.51 & 1.62 & 2.30 & 1.51 & 1.70
& \multirow{2}{*}{$4.22 \pm 0.08$} \\

& SPK-SIM $\uparrow$
& 0.678
& 0.780 & 0.716 & 0.674 & 0.644 & 0.615 & 0.588 & 0.560
& \\

\noalign{\vskip 0.6pt}
\cdashline{1-11}[0.3pt/1.5pt]
\noalign{\vskip 0.6pt}

\multirow{2}{*}{MOSS-TTS-v1.5~\cite{gong2026mossttstechnicalreport}}
& WER $\downarrow$
& 1.77
& 1.33 & 1.25 & 1.42 & 1.57 & 1.49 & 1.36 & 1.34
& \multirow{2}{*}{$4.35 \pm 0.08$} \\

& SPK-SIM $\uparrow$
& 0.709
& 0.757 & 0.726 & 0.707 & 0.698 & 0.690 & 0.687 & 0.679
& \\

\specialrule{0.8pt}{1pt}{1pt}

\multirow{2}{*}{VoXtream2~\cite{torgashov2026voxtream2fullstreamttsdynamic}}
& WER $\downarrow$
& 8.63
& 3.60 & 9.73 & 10.20 & 10.34 & 10.25 & 8.47 & 9.79
& \multirow{2}{*}{$4.03 \pm 0.08$} \\

& SPK-SIM $\uparrow$
& 0.659
& 0.734 & 0.695 & 0.660 & 0.637 & 0.627 & 0.618 & 0.595
& \\

\noalign{\vskip 0.6pt}
\cdashline{1-11}[0.3pt/1.5pt]
\noalign{\vskip 0.6pt}

\multirow{2}{*}{\textbf{Ours}}
& WER $\downarrow$
& 3.58
& 3.42 & 3.42 & 3.84 & 3.64 & 3.63 & 3.39 & 3.96
& \multirow{2}{*}{$4.03 \pm 0.09$} \\

& SPK-SIM $\uparrow$
& 0.759
& 0.770 & 0.766 & 0.763 & 0.761 & 0.760 & 0.756 & 0.754
& \\

\bottomrule
\end{tabular*}
\vskip -10pt

\end{table*}

%% file: sections/3_experiments.tex
\section{Experiments}
\label{sec:experiments}

\subsection{Experimental Setups}
\label{subsec:experimental_setups}

All VoXtream2-based~\cite{torgashov2026voxtream2fullstreamttsdynamic} experiments use the official implementation and pretrained checkpoint. Our method requires no additional training. We follow the same LibriSpeech-Long~\cite{pmlr-v267-park25k} evaluation protocol and objective metrics as in Section~\ref{subsec:longform}. In addition to VoXtream2 and ours, we compare against Qwen3-TTS-12Hz-1.7B-Base~\cite{hu2026qwen3ttstechnicalreport} and MOSS-TTS-v1.5~\cite{gong2026mossttstechnicalreport}, recent TTS models capable of long-form synthesis. Both baselines use long-form training, whereas ours directly extends pretrained VoXtream2 without additional training.

All models receive the same acoustic prompt and target text; when required, the corresponding reference text is also provided. We further evaluate Mimi codec reconstruction~\cite{defossez2024moshispeechtextfoundationmodel} to quantify degradation introduced by encoding and decoding. All audio is resampled to 16kHz for evaluation. In addition to WER and SPK-SIM, we evaluate naturalness and overall speech quality using mean opinion score (MOS). For MOS evaluation, 158 listeners rate overall naturalness and audio quality on a five-point scale, yielding 325 ratings per model.

\input{tables/Table2}

\subsection{Results}
\label{subsec:results}

\subsubsection{Model Comparison}
\label{subsubsec:model_comparison}

Table~\ref{tab:main_results} compares long-form synthesis performance.
Mimi reconstruction largely preserves WER (1.86 to 1.95) but noticeably reduces SPK-SIM from 0.832 to 0.754.
Since VoXtream2 generates speech through the Mimi codec, its reconstruction MOS of 4.10 also provides a practical upper bound on perceptual quality under this codec.
Our method substantially improves VoXtream2, reducing overall WER from 8.63\% to 3.58\% and increasing SPK-SIM from 0.659 to 0.759.
The improvement is particularly evident over long generation: while VoXtream2 progressively degrades with generation length, our method maintains WER below 4.0\% and nearly stable SPK-SIM across 30s segments.
Nevertheless, even within its stable operating range, VoXtream2 already exhibits lower pronunciation accuracy (3.60\%), and our training-free extension therefore does not reach the WER of Qwen3-TTS or MOSS-TTS.
Our method achieves the same MOS of 4.03 as VoXtream2, close to the Mimi reconstruction upper bound of 4.10, indicating that the improved long-form robustness does not compromise perceptual quality.

Interestingly, speaker similarity degradation with generation length is not unique to VoXtream2. Qwen3-TTS and MOSS-TTS, despite being trained with long-form data, also exhibit decreasing SPK-SIM toward later segments. Qwen3-TTS, for example, drops from 0.780 to 0.560, whereas our method remains substantially more stable, from 0.770 to 0.754. This suggests that maintaining speaker identity over extended generation remains a broader challenge even for models explicitly trained for long-form synthesis.

\subsubsection{Ablation Studies}
\label{subsubsec:ablation}

Table~\ref{tab:ablation} analyzes context window restriction and acoustic prompt attention bias.
Reducing the Temporal Transformer window from $W=625$ improves WER, with $W=375$ achieving the best value of 4.82, whereas further reducing it to $W=250$ degrades performance.
Despite more frequent cache resets, $W=375$ remains real-time on an NVIDIA RTX 3090, with real-time factor (RTF) increasing only from 0.415 at $W=625$ to 0.433.
With $W=375$, increasing the constant bias improves speaker similarity; $b=2.0$ achieves the highest SPK-SIM of 0.766, while $b=1.5$ yields the lowest WER of 3.58 with a comparable SPK-SIM of 0.759.
SpotLight-based adaptive steering~\cite{venkateswaran-contractor-2026-spotlight} improves WER over window restriction alone but remains less effective than constant bias for speaker similarity.
We therefore use $W=375$ and a constant bias of $b=1.5$ for our experiments.

%% file: tables/Table2.tex
\begin{table}[t]
\centering
\caption{
Ablation study on window size and attention bias.
$W$, $b$, and $\psi_{\mathrm{target}}$ denote the window size, bias strength, and target attention mass, respectively.
}
\label{tab:ablation}
\vspace{1mm}

\small
\renewcommand{\arraystretch}{1.0}

\begin{tabular}{lccc}
\toprule
\textbf{Setting}
& \textbf{Value}
& \textbf{WER (\%) $\downarrow$}
& \textbf{SPK-SIM $\uparrow$} \\
\midrule

\rowcolor{gray!15}
\multicolumn{4}{l}{\textit{Window size (No bias)}} \\

\addlinespace[1pt]

Default
& $W=625$ & 8.63 & 0.659 \\

\noalign{\vskip 0.6pt}
\cdashline{1-4}[0.3pt/1.5pt]
\noalign{\vskip 0.6pt}

\multirow{3}{*}{Window}
& $W=500$ & 5.42 & 0.668 \\
& $W=375$ & 4.82 & 0.683 \\
& $W=250$ & 6.51 & 0.703 \\

\midrule

\rowcolor{gray!15}
\multicolumn{4}{l}{\textit{Attention bias ($W=375$)}} \\

\addlinespace[1pt]

No Bias
& -- & 4.82 & 0.683 \\

\noalign{\vskip 0.6pt}
\cdashline{1-4}[0.3pt/1.5pt]
\noalign{\vskip 0.6pt}

\multirow{4}{*}{Constant}
& $b=0.5$ & 4.01 & 0.721 \\
& $b=1.0$ & 3.65 & 0.744 \\
& $b=1.5$ & 3.58 & 0.759 \\
& $b=2.0$ & 3.65 & 0.766 \\

\noalign{\vskip 0.6pt}
\cdashline{1-4}[0.3pt/1.5pt]
\noalign{\vskip 0.6pt}

\multirow{5}{*}{Adaptive}
& $\psi_{\mathrm{target}}=0.1$ & 3.73 & 0.679 \\
& $\psi_{\mathrm{target}}=0.2$ & 3.69 & 0.682 \\
& $\psi_{\mathrm{target}}=0.3$ & 4.02 & 0.686 \\
& $\psi_{\mathrm{target}}=0.4$ & 3.90 & 0.692 \\
& $\psi_{\mathrm{target}}=0.5$ & 3.84 & 0.701 \\

\bottomrule
\end{tabular}
\vskip -14pt
\end{table}

%% file: sections/4_conclusion.tex
\section{Conclusion}
In this paper, we proposed a training-free approach for long-form generation with full-stream TTS, based on a systematic analysis of VoXtream2. By shrinking the attention window to keep the model within the positions encountered during training and applying a bias to the attention logits of the acoustic prompt, we preserved both pronunciation accuracy and speaker similarity throughout long-form generation while retaining the naturalness of the original model.

%% file: main.bib
@inproceedings{
xiao2024efficient,
title={Efficient Streaming Language Models with Attention Sinks},
author={Guangxuan Xiao and Yuandong Tian and Beidi Chen and Song Han and Mike Lewis},
booktitle={The Twelfth International Conference on Learning Representations},
year={2024},
url={https://openreview.net/forum?id=NG7sS51zVF}
}

@misc{du2024cosyvoice2scalablestreaming,
      title={CosyVoice 2: Scalable Streaming Speech Synthesis with Large Language Models}, 
      author={Zhihao Du and Yuxuan Wang and Qian Chen and Xian Shi and Xiang Lv and Tianyu Zhao and Zhifu Gao and Yexin Yang and Changfeng Gao and Hui Wang and Fan Yu and Huadai Liu and Zhengyan Sheng and Yue Gu and Chong Deng and Wen Wang and Shiliang Zhang and Zhijie Yan and Jingren Zhou},
      year={2024},
      eprint={2412.10117},
      archivePrefix={arXiv},
      primaryClass={cs.SD},
      url={https://arxiv.org/abs/2412.10117}, 
}

@misc{bai2025speakstreamstreamingtexttospeechinterleaved,
      title={SpeakStream: Streaming Text-to-Speech with Interleaved Data}, 
      author={Richard He Bai and Zijin Gu and Tatiana Likhomanenko and Navdeep Jaitly},
      year={2025},
      eprint={2505.19206},
      archivePrefix={arXiv},
      primaryClass={cs.CL},
      url={https://arxiv.org/abs/2505.19206}, 
}

@misc{torgashov2026voxtream2fullstreamttsdynamic,
      title={VoXtream2: Full-stream TTS with dynamic speaking rate control}, 
      author={Nikita Torgashov and Gustav Eje Henter and Gabriel Skantze},
      year={2026},
      eprint={2603.13518},
      archivePrefix={arXiv},
      primaryClass={eess.AS},
      url={https://arxiv.org/abs/2603.13518}, 
}

@InProceedings{pmlr-v267-park25k,
  title = 	 {Long-Form Speech Generation with Spoken Language Models},
  author =       {Park, Se Jin and Salazar, Julian and Jansen, Aren and Kinoshita, Keisuke and Ro, Yong Man and Skerry-Ryan, Rj},
  booktitle = 	 {Proceedings of the 42nd International Conference on Machine Learning},
  pages = 	 {48245--48261},
  year = 	 {2025},
  editor = 	 {Singh, Aarti and Fazel, Maryam and Hsu, Daniel and Lacoste-Julien, Simon and Berkenkamp, Felix and Maharaj, Tegan and Wagstaff, Kiri and Zhu, Jerry},
  volume = 	 {267},
  series = 	 {Proceedings of Machine Learning Research},
  month = 	 {13--19 Jul},
  publisher =    {PMLR},
  url = 	 {https://proceedings.mlr.press/v267/park25k.html}
}

@misc{ghosh2026magpiettslfinferencetimelongformspeech,
      title={MagpieTTS-LF: Inference-Time Long-Form Speech Generation Without Training on Long-Form data}, 
      author={Subhankar Ghosh and Jason Li and Paarth Neekhara and Shehzeen Hussain and Ryan Langman and Xuesong Yang and Roy Fejgin},
      year={2026},
      eprint={2606.18485},
      archivePrefix={arXiv},
      primaryClass={cs.SD},
      url={https://arxiv.org/abs/2606.18485}, 
}

@misc{liu2026prosodicboundaryawarestreaminggeneration,
      title={Prosodic Boundary-Aware Streaming Generation for LLM-Based TTS with Streaming Text Input}, 
      author={Changsong Liu and Tianrui Wang and Ye Ni and Yizhou Peng and Eng Siong Chng},
      year={2026},
      eprint={2603.06444},
      archivePrefix={arXiv},
      primaryClass={cs.SD},
      url={https://arxiv.org/abs/2603.06444}, 
}

@misc{defossez2024moshispeechtextfoundationmodel,
      title={Moshi: a speech-text foundation model for real-time dialogue}, 
      author={Alexandre Défossez and Laurent Mazaré and Manu Orsini and Amélie Royer and Patrick Pérez and Hervé Jégou and Edouard Grave and Neil Zeghidour},
      year={2024},
      eprint={2410.00037},
      archivePrefix={arXiv},
      primaryClass={eess.AS},
      url={https://arxiv.org/abs/2410.00037}, 
}

@InProceedings{pmlr-v202-radford23a,
  title = 	 {Robust Speech Recognition via Large-Scale Weak Supervision},
  author =       {Radford, Alec and Kim, Jong Wook and Xu, Tao and Brockman, Greg and Mcleavey, Christine and Sutskever, Ilya},
  booktitle = 	 {Proceedings of the 40th International Conference on Machine Learning},
  pages = 	 {28492--28518},
  year = 	 {2023},
  editor = 	 {Krause, Andreas and Brunskill, Emma and Cho, Kyunghyun and Engelhardt, Barbara and Sabato, Sivan and Scarlett, Jonathan},
  volume = 	 {202},
  series = 	 {Proceedings of Machine Learning Research},
  month = 	 {23--29 Jul},
  publisher =    {PMLR},
  url = 	 {https://proceedings.mlr.press/v202/radford23a.html}
}

@inproceedings{desplanques20_interspeech,
  title     = {{ECAPA-TDNN: Emphasized Channel Attention, Propagation and Aggregation in TDNN Based Speaker Verification}},
  author    = {Brecht Desplanques and Jenthe Thienpondt and Kris Demuynck},
  year      = {2020},
  booktitle = {{Interspeech 2020}},
  pages     = {3830--3834},
  doi       = {10.21437/Interspeech.2020-2650},
  issn      = {2958-1796},
}

@misc{shi2026qwen3asrtechnicalreport,
      title={Qwen3-ASR Technical Report}, 
      author={Xian Shi and Xiong Wang and Zhifang Guo and Yongqi Wang and Pei Zhang and Xinyu Zhang and Zishan Guo and Hongkun Hao and Yu Xi and Baosong Yang and Jin Xu and Jingren Zhou and Junyang Lin},
      year={2026},
      eprint={2601.21337},
      archivePrefix={arXiv},
      primaryClass={cs.CL},
      url={https://arxiv.org/abs/2601.21337}, 
}

@misc{hu2026qwen3ttstechnicalreport,
      title={Qwen3-TTS Technical Report}, 
      author={Hangrui Hu and Xinfa Zhu and Ting He and Dake Guo and Bin Zhang and Xiong Wang and Zhifang Guo and Ziyue Jiang and Hongkun Hao and Zishan Guo and Xinyu Zhang and Pei Zhang and Baosong Yang and Jin Xu and Jingren Zhou and Junyang Lin},
      year={2026},
      eprint={2601.15621},
      archivePrefix={arXiv},
      primaryClass={cs.SD},
      url={https://arxiv.org/abs/2601.15621}, 
}

@misc{gong2026mossttstechnicalreport,
      title={MOSS-TTS Technical Report}, 
      author={Yitian Gong and Botian Jiang and Yiwei Zhao and Yucheng Yuan and Kuangwei Chen and Yaozhou Jiang and Cheng Chang and Dong Hong and Mingshu Chen and Ruixiao Li and Yiyang Zhang and Yang Gao and Hanfu Chen and Ke Chen and Songlin Wang and Xiaogui Yang and Yuqian Zhang and Kexin Huang and ZhengYuan Lin and Kang Yu and Ziqi Chen and Jin Wang and Zhaoye Fei and Qinyuan Cheng and Shimin Li and Xipeng Qiu},
      year={2026},
      eprint={2603.18090},
      archivePrefix={arXiv},
      primaryClass={cs.SD},
      url={https://arxiv.org/abs/2603.18090}, 
}

@inproceedings{venkateswaran-contractor-2026-spotlight,
    title = "Spotlight Your Instructions: Instruction-following with Dynamic Attention Steering",
    author = "Venkateswaran, Praveen  and
      Contractor, Danish",
    editor = "Demberg, Vera  and
      Inui, Kentaro  and
      Marquez, Llu{\'i}s",
    booktitle = "Proceedings of the 19th Conference of the {E}uropean Chapter of the {A}ssociation for {C}omputational {L}inguistics (Volume 1: Long Papers)",
    month = mar,
    year = "2026",
    address = "Rabat, Morocco",
    publisher = "Association for Computational Linguistics",
    url = "https://aclanthology.org/2026.eacl-long.174/",
    doi = "10.18653/v1/2026.eacl-long.174",
    pages = "3752--3770",
    ISBN = "979-8-89176-380-7"
}

@misc{guardieiro2026instructionfollowingprincipledboosting,
      title={Instruction Following by Principled Boosting Attention of Large Language Models}, 
      author={Vitoria Guardieiro and Avishree Khare and Adam Stein and Eric Wong},
      year={2026},
      eprint={2506.13734},
      archivePrefix={arXiv},
      primaryClass={cs.CL},
      url={https://arxiv.org/abs/2506.13734}, 
}
